
\input phyzzx
\baselineskip=10pt
\overfullrule=0pt
\def\co{cohomology\ }
\def\G{${G\over G}\ $}

\def\ch#1#2{\chi_{#1}^{#2}}

\def\rh#1#2{\rho_{#1}^{#2}}
\def\jt#1#2{{J^{(tot)}}_{#1}^{#2}}
\def\y{|phys>}
\def\w{\omega}
\def\t{\tilde}
 \def\pa{\partial}
\def \f{f^a{}_{bc}}
\def \fcr{\f \ch {} b \rh {} c }
\def \jg{J^{(gh)}}

\def \pa{\partial}

\def\S#1{$SL({#1},R)$}
\def\SOS {$SL(2,R)\over SL(2,R)$}
\def\SOU {$SL(2,R)\over U(1)$}
\def\A {$A_{N-1}^{(1)}$}

\def\W{$W_N\ $ }
\def\G{${G\over G}\ $}
\def\GH{${G\over H}\ $}
\def\tGH{twisted\ ${G\over H}\ $}

 \def\t{\tilde}
 \def\pa{\partial}

\def\bJ{\beta_J}
\def\bI{\beta_I}
\def\bP{\beta^+}
\def\bM{\beta^-}

\def\gJ{\gamma_J}
\def\gI{\gamma_I}
\def\gP{\gamma^{+}}
\def\gM{\gamma^{-}}

\def\pJ{\phi_J}
\def\pI{\phi_I}
\def\pP{\phi^+}
\def\pM{\phi^-}

\def\dpJ{\partial\phi_J}
\def\dpI{\partial\phi_I}
\def\dpP{\partial\phi^+}
\def\dpM{\partial\phi^-}

\def\cP{\chi^+}
\def\c0{\chi^0}
\def\cM{\chi^-}

\def\rP{\rho^+}
\def\r0{\rho^0}
\def\rM{\rho^-}

\def\hlf{{1 \over 2}}
\def\sqh{{\sqrt{t \over 2}}}

\def\sq2{{\sqrt 2}}
\def\sqhnoi{{1 \over \sqrt {2t}}}
\def\d{\partial}

\def\wz{{z -\w }}
\def\QB{Q_{BRST}}
\def\QBr{{\QB}^{(red)}}
\def\Qt{Q_{tr}^{(red)}}
\def\intz{\oint{dz\over{2\pi i}}}

\def\cmp#1{{\it Comm. Math. Phys.} {\bf #1}}

\def\pl#1{{\it Phys. Lett.} {\bf #1B}}

\def\prd#1{{\it Phys. Rev.} {\bf D#1}}

\def\np#1{{\it Nucl. Phys.} {\bf B#1}}

\def\jmath#1{{\it J. Math. Phys.} {\bf #1}}
\def\mpl#1{{\it Mod. Phys. Lett.}{\bf A#1}}

\REF\dvv{R. Dijkgraaf, E. Verlinde, and H. Verlinde,
\np {352} (1991) 59;
``Notes On Topological String Theory And 2D Quantum Gravity,'' Princeton
preprint PUPT-1217 (1990).}
\REF\us  {O. Aharony,O. Ganor N. Sochen  J. Sonnenschein and S.
Yankielowicz ,``Physical states in the \G models and two dimensional
gravity", TAUP- 1947-92 April 1992 to appear in \np{}. }
\REF\uss  { O. Aharony, J. Sonnenschein and S.
Yankielowicz ,``  \G models and \W strings" ,
 \pl {289}  (1992) 309.}
\REF\usss  {O. Aharony,O. Ganor  J. Sonnenschein and S.
Yankielowicz ,``On the \tGH topological models",
TAUP- 1990-92 August 1992 to appear in \np{}.}
\REF\DDK{F. David \mpl {3}  (1988) 1651;
\break J. Distler and H. Kawai \np
{321}
(1989) 509.}
\REF\Wbh{ E. Witten \prd {44} (1991) 314.}
\REF\MuVa{ S. Mukhi and C. Vafa `` Two Dimensional Black hole as a
Topological Coset model of $c=1$ String Theory",HUTP-93/A002,
TIFR/TH/93-01.}
\REF\BF{D. Bernard  and G. Felder \cmp {127}  (1991)  145.}
\REF\BaRS{K. Bardacki, E. Rabinovici, and B. Serin \np {299} (1988) 151.}
\REF\GK{ K. Gawedzki and A. Kupianen , \pl {215} (1988) 119, \np
     {320} (1989)649.}
\REF\KS{D. Karabali and H. J. Schnitzer, \np {329} (1990) 625.}
\REF\BMP{P. Bouwknegt, J. McCarthy and  K. Pilch Cern Preprint
TH-6162/91}
\REF\BMPN{P. Bouwknegt, J. McCarthy and  K. Pilch \pl {234} (1990) 297,
\cmp {131} (1990) 125.}

\REF\IsKa{ H. Ishikawa and M. Kato `` Equivalence of BRST \co for
2-d black hole and c=1 Liouville theory" UT-Komaba/92-11 Nov. 1992.}
 \REF\Wgr{ E. Witten  \np {377} (1992) 55.}
\REF\SY{M. Spiegelglas and S. Yankielowicz
`` \G Topological Field Theories by
Cosetting $G_k$" TAUP-1934 .
;``Fusion Rules As Amplitudes in $G/G$ Theories,'' Technion PH- 35-90
to appear in \np{}}
\REF\Wgg{E. Witten,
\cmp 144 (1992) 189.}
\REF\Comment{ For a comment about a possible
difficulty in case of non-compact
groups see ref. [\uss]}

\REF\KaSu{ Y. Kazama and H. Suzuki, \np {321} (1989).}
\REF\Wak{M. Wakimoto,   \cmp   {104}  (1989) 605.}

\REF\FeFr{B. Feigin and E. Frenkel \pl {246} (1990) 75.}
\REF\BLNW{ M. Bershadsky, W. Lerche, D. Nemeschansky and N. P. Warner
``A BRST operator for  non-critical W strings" CERN-TH  6582/92.}
\REF\NaSu{ T. Nakatsu and Y. Sugawara
`` Topological gauged WZW  models and 2 D gravity" Tokyo preprint  UT-598
``BRST fixed points  and Topological Conformal Symmetry" UT-599.}
\REF\HuYu { L. H. Hu and M. Yu,
`` On BRST cohomology of SL(2)/SL(2) gauged WZWN
models" Academia Sinica preprint AS-ITP-92-32.}
\REF\Wgga{E. Witten,  \np {371} (1992) 191.}
\rightline{TAUP- 2032-93}
\date{February 1993}
\titlepage
\vskip 1cm
\title{c=1 String Theory as a Topological \G Model}
\author {O. Aharony, O. Ganor , J. Sonnenschein and S. Yankielowicz
\footnote{\dagger}{Work supported
in part by the US-Israel Binational Science
Foundation and the Israel Academy of Sciences.}}
\address{ School of Physics and Astronomy\break
Beverly and Raymond Sackler \break
Department of Exact Sciences\break
Tel-Aviv University\break
Ramat Aviv Tel-Aviv, 69987, Israel}
\abstract{
The  physical states on the free
field Fock space   of the \SOS\  model at
any level are
computed. Using a similarity transformation on $\QB$, the cohomology
of the latter is
mapped into a direct sum of simpler cohomologies.
We show a one to one correspondence
between the states of the $k=-1$ model and those of
the $c=1$ string model. A full
equivalence between the \SOS\ and \SOU\  models at the level of their
 Fock space cohomologies is found.}\endpage
A deeper understanding of the relation between topological theories,
N=2 theories and string theories is one of the more challenging
problems of string theory in recent years. The question
 ``Is the bosonic string topological?"
which was raised in ref. [\dvv] should be stated in a broader context,
namely,  to what extent do string theories
admit a topological description.  Unravelling this relation
  may lead  not only  to a different formulation of string theory
but also  to better computational tools.

In previous publications\refmark{\us,\uss,\usss} we have put
forward the equivalence between twisted topological \G models
 of \A\ at level   $k={p\over q}-N$  and the $(p,q)$
$W_N$ minimal models coupled to $W_N$ gravity.
We have also demonstrated that the \co ring is the same  for
twisted \GH topological models  with $rank\ G=rank\
H$. In this paper we investigate the twisted
$SL(2,R)\over SL(2,R)$ theory at level $k=-1$, derive the space of
physical states  and establish its equivalence to  that of the $c=1$
string theory coupled to gravity.\refmark\DDK\
First indications towards this
equivalence were already given  in ref. [\us]. In particular we
identify the tachyon operators as well as the ground ring
generators. We further discuss the generalization of our methods to
other \S 2 levels which correspond to $c<1$ non-critical string
theories, as well as to the twisted ${SL(2,R)\over U(1)}$ case
which is related to the 2D black hole.\refmark\Wbh\
The \SOU\ topological model at level $k=-3$
is studied at length in a  recent paper by
Mukhi and Vafa.\refmark\MuVa\  In their paper the
equivalence of this topological
model and the $c=1$ string model is analyzed.
In the present paper we show that  the
Fock space \co of  all these models at different
levels are essentially the
same. However, unlike the $c=1$ case,  for the $c<1$ models one has
to  employ  a further reduction.\refmark\BF

The paper is organized as follows:
In section 1 the basic building
blocks are presented namely the free field bosonization,
OPEs, and  the BRST operator.
A special map is invoked to translate the cohomology into
a sum of simpler cohomologies. The complete BRST
cohomology is then extracted. In section
2 a comparison between the
Fock space states of non-critical string models  and that of
the various \SOS\ models is made.
In particular the ground ring as well as the
tachyonic branches of the $c=1$ model
are  identified in the \G picture. The full
equivalence is achieved only after allowing arbitrary powers of certain
current components.  Section 3 is
devoted to a brief description of the application of
the procedure developed for the \SOS\ models to the  \SOU\ models.
We then
summarize, briefly  compare to some
recent works and discuss some open questions.
An appendix is devoted to the
 derivation of the transformation of the BRST
charge and its corresponding cohomology.
 \section{ BRST Cohomology}
 The   \G topological
 model\refmark{\SY,\Wgg} is constructed by gauging
the anomaly free diagonal $G$ group of the WZW model for the group $G$.
 The quantum action  of the
 model was shown to be composed of three decoupled
parts:\refmark{\BaRS,\GK,\KS}
  $S_k(g)-$ a $WZW$ model of level
  $k$ with $g\in G$, $S_{-(k+2C_G)}(h)-$ a
$WZW$ model of level
$-(k+2C_G)$ with $h\in G$,
 and a  dimension $(1,0)$ system of anticommuting ghosts $\rho$ and $\chi$
  in the adjoint representation of the group. The action, thus,
reads\refmark\Comment
 $$S_k(g,h,\rho,\chi) =S_k(g) +S_{-(k+2C_G)}(h) -i\int d^2z
Tr[\bar\rho \pa \bar\chi + \rho\bar\pa\chi], \eqn\mishwzwh$$
 where  $C_G$ is the second Casimir of the
adjoint representation.

Invariance of each of the three terms under holomorphic
$G$ transformations implies that there are three  Kac-Moody currents
$J(z)= g^{-1}\pa g$, $I(z)= h^{-1}\pa h$ and ${\jg}^a=\fcr$
of levels  $k$,$-(k+2c_G)$ and $2c_G$
respectively. The twisted theory for $G=SL(2,R)$ is
obtained by replacing the energy
momentum tensor $T$ of this theory with $\t T =T+\pa\jt {} 0$, where
$\jt {} {}$
is the sum of the holomorphic currents from all sectors (and, therefore,
is at level $k=0$).

The space of physical states of the twisted \G models
which correspond to minimal matter models coupled to gravity was
extracted\refmark{\us,\uss,\usss} using a spectral sequence
decomposition approach\refmark\BMP. In that formulation we had to
use a particular Wakimoto realization of the matter ($J$) and ``gauge"
($I$) sectors\refmark{\us}. There are two possible bosonizations of the
$SL(2,R)$ current algebra, which are related by the automorphism
$J^+ \leftrightarrow J^-, J^0 \leftrightarrow -J^0$. Let us denote the
bosonization given in eqn. (2) as the $(+)$ bosonization,
 and the  one related to it by the automorphism
as the $(-)$ bosonization. In [\us] we used a $(+)$ bosonization in the
$J$ sector and a $(-)$ bosonization in the $I$ sector. We shall call this
the $(+,-)$ bosonization of the theory.
The physical states were associated with the
cohomology ring on the space of Kac-Moody irreducible
representations via a projection due to Bernard and Felder\refmark\BF.
Note that  the  $(+,-)$ bosonization lacks an $SL(2,R)$
invariant vacuum. For the 2D gravity coupled to
minimal matter this was not a problem,
since  after the Bernard-Felder cohomology is performed
both bosonizations
give an irreducible representation of the
Kac-Moody algebra,
(The $SL(2,R)$ invariant vacuum is restored, since the only problem
arises from  $L_{-1}|vacuum> \neq 0$. However, this  is  a descendant
of a null in any case
and therefore is zero after the Bernard-Felder reduction.)
We now introduce the following
bosonization\refmark\Wak\ (with normal ordering
assumed everywhere) :
$$\eqalign{J^+ = \bJ\ \ \ \ I^+ =& \bI \cr
 J^0 = \bJ\gJ + i\sqrt{t\over2}\dpJ \ \ \ \  I^0 =& \bI\gI +
\sqrt{t\over2}\dpI\cr
 J^- = -\bJ\gJ^2 - i\sqrt{2t}\gJ\dpJ - (t-2)\d\gJ\ \ \ \
 I^- =& -\bI\gI^2 -  \sqrt{2t}\gI\dpI + (t+2)\d\gI\cr}\eqn\mishbos$$
where  $\bJ,\ \gJ$ and $\bI,\ \gI$ form bosonic $(1,0)$
systems,    $\phi_J $ and  $\phi_I $ are free  scalar fields with
background charges of ${-i\over \sqrt{2t}}$ and ${1\over \sqrt{2t}}$
respectively, and $t=k+2$. It will turn out to be convenient
to use the following linear combinations of  the free fields of the
$J$ and $I$ sector :
$$\bP=\bJ+\bI,\qquad\gP=\hlf(\gJ+\gI),\qquad\pP=\sqhnoi(\pJ+i\pI)$$
$$\bM=\bJ-\bI,\qquad\gM=\hlf(\gJ-\gI),
\qquad\pM=\sqh(\pJ-i\pI).\eqn\mishPM$$
They obey  the following  OPEs:
$$\gP(z)\bP(\w)=\gM(z)\bM(\w)={1\over\wz} + \dots$$
$$\dpP(z)\pM(\w)=\dpM(z)\pP(\w)={-1\over\wz} + \dots$$
$$\cP(z)\rM(\w)=\cM(z)\rP(\w)=2\c0(z)\r0(w)={1\over\wz}+\dots\eqn\mishOP$$

In terms of this free field realization $\jt {} {0}   =J^0 +I^0 +{\jg}^0$
and the  energy momentum tensor   now read
$${J^0}^{(total)} =\bP\gP+\bM\gM +i\dpM+\cP\rM-\cM\rP.\eqn\mishJ$$
$$T^{(total)} =-\dpP\dpM -i\pa^2\pP -\bP\pa\gP -\bM\pa\gM
-\rP\d\cM-\rM\d\cP-2\r0\d\c0. \eqn\mishT$$
The twisted energy momentum tensor with which we define our theory is
therefore given by :
$$\t T =(-\dpP\dpM -i\pa^2\pP +i\pa^2\pM -2\rM\d\cP+\cP\d\rM)
+\gP\pa\bP +\gM\pa\bM
-\cM\d\rP-2\r0\d\c0. \eqn\mishTTw$$
As noted in [\us], if we identify $\pJ$ with $X$ of the $c\leq 1$
Liouville models, $\pI$ with the Liouville field, $\cP$ with $c$
and $\rM$ with $b$, the first part  (in parenthesis)
of the twisted
energy momentum tensor, as depicted, in \mishTTw\
equals exactly the energy tensor of the
Liouville theory at $c={-6t^2+13t-6 \over t}$. The rest of $\t T$
is composed of two pairs of $(1,0)$ bosons and $(1,0)$ fermions which
could be treated as additional "topological sectors". Note that both
$t$ and $1\over t$ give an energy momentum tensor which can be identified
with the Liouville theory at the same $c$. Moreover,  we can also identify
the Liouville theory at $c(t)$ with the \SOS\ theory at $-t$ if we change our
identifications so that $\pJ$ is identified with the Liouville field
and  $\pI$ with the matter field.
The BRST charge takes the form
$$\eqalign {\QB=\intz(&\cM(J^+ + I^+)+2\c0(J^0 + I^0) +\cP(J^- + I^-)\cr
    &+2(\cP\cM\r0 + \cM\c0\rP + \c0\cP\rM)).\cr } \eqn\mishQ$$
Both $\t T$ (with which we will work from here on)
and $\jt {} a$ are $Q$ exact : $\t T = \{Q, G\}$ for
$G= \rh {} - (J^+
- I^+) + 2\rh {} 0 (J^0 - I^0) + \rh {} +(J^- - I^-) +  \pa\rh {} 0$
and $\jt {} a = \{Q, \rh {} a\}$.
Since  both  $\jt n a $ and $L_n$ are  $Q$ exact
 it follows that
$$L_0\y =  0 \qquad \jt 0 0 \y =  0. \eqn\mishL$$
Let us select a subspace of the
space of physical states on which we further
impose $\rh 0 0 |phys>=0$.
On this subspace $\QB$  is reduced to $\QBr$ which
does not include $\rh 0 0$ and $\ch 0 0 $ as
follows from the decomposition
$$\QB = \c0_0 {J^0_0}^{(total)} + \r0_0 M + \QBr   \eqn\mishah$$
where
$M = 2\sum_n\cP_{-n}\cM_n$.
Once we compute the $\QBr$ cohomology, it will be easy to obtain the full
$\QB$ cohomology, since the representatives (modulo $\QB$ exact states)
of the physical states would be :\refmark\BMP
$$N|\Psi> + \c0_0|\Psi> + |\Psi'>   \eqn\mishnba$$
where $N = \sum_n\cP_{-n}\gP_n$, and
 $|\Psi>$ and $|\Psi'>$ are representatives of the $\QBr$ cohomology.
 A direct computation of the \co of $\QBr$
in the bosonization of eqn. \mishbos\ is not an easy task
due to the  appearance of cubic and quartic terms.
    We thus follow a different route where the  latter is
mapped\refmark\IsKa\
 into a nilpotent
operator   which is a sum of  anti-commuting terms
acting on different sectors of the theory.  Therefore, the
corresponding \co is a direct sum of simpler cohomologies.
Let us  define  the  dimension (0,0) operators of zero ghost number
$$\eqalign{R&=\intz(\cP\rP\gM\gM + 2\cP\r0\gP - \cP\rP\gP\gP)\cr
P&=-\intz'(i\pP(\bP\gP+\bM\gM+\cP\rM-\cM\rP))',\cr}\eqn\mishPR$$
where $\intz'$ means that the zero modes of $\pP$ were excluded.
We then use these operators to transform $\QBr$
to the desired form (for details see the appendix) in the
following way $$e^{-P}e^R \QBr e^{-R}e^P = \Qt\eqn\mishQpri$$
with
$$ \eqalign{ \Qt &= \intz[\cM\bP+2i\c0\dpM
-2t\cP\d\gM-2t\pP_0\cP\gM]\cr
&= 2\sum_{n\neq 0}\c0_{-n}\pM_n
+\sum_n(\cM_{-n}\bP_n-2t(\pP_0-n-1)\cP_{-n}\gM_n).\cr}\eqn\mishQtr$$
The mode expansions are relative to the vacuum of the
twisted theory (i.e. $\gamma(z) =\sum_n\gamma_n z^{-(n+1)}$).
{}From \mishQpri\ it follows that the cohomologies of $\QBr$ and of $\Qt$
are isomorphic,
namely, for every state $|\Phi_0>$ in the cohomology of $\Qt$,
the state $|\Psi>=e^{-R}e^P|\Phi_0>$ is in the cohomology of $\QBr$ and
vice versa.

  On the following  direct sum of  Fock spaces
$$\bigoplus_{n\neq 0}F(\c0_{-n},\r0_n,\pM_{-n},\pP_n)
  \bigoplus_n F(\cM_{-n},\rP_n,\gP_{-n},\bP_n)
  \bigoplus_n F(\cP_{-n},\rM_n,\bM_{-n},\gM_n)\eqn\mishFo$$
the first term is subjected to the action of the first term in eqn.
\mishQtr, and  similarly for the second and third terms.
  It is thus  apparent that $\Qt$ indeed decomposes into a sum of
anti-commuting terms which act on separate Fock spaces and, therefore,
that the  \co ring  is a direct sum of smaller ones.
In  the first and second parts of the Fock space, the cohomology ring
includes only a single state which is the corresponding   vacuum state. In
the third part one finds states, in addition to the vacuum, for
 $\pP_0=n +1$ when $n$ is an integer.
The vacuum  $|0>_{phys}$
corresponds to the twisted energy-momentum tensor $\t T
=T+\pa\jt {} 0$.
It is related to the $SL(2,C)$ invariant vacuum $|0>_{SL(2,C)}$
in the following
way: $|0>_{phys}=\ch 1 +  |0>_{SL(2,C)}$, and it is annihilated by the
following zero modes: $\ch 0 - |0>_{phys}=\gamma_0^{\pm}
|0>_{phys} =\rh 0 - |0>_{phys} = 0.$
The only  zero modes acting on the vacuum are, therefore,
$\ch 0 + , \rh 0 + $ and $
 \beta_0^{\pm}$.  This is a lowest weight vacuum for both \S 2 algebras.

 In the  case   $\pP_0=-n+1,\ n> 0$, the
  cohomology  of $\Qt$ is in the
span of the states $${\rM_{-n}}^s{\gM_{-n}}^r|\pM_0,\pP_0>$$
where $s=0,1$ and $r=0,1,\dots$ and
$|\pM_0,\pP_0>$ is the vacuum state with momenta $\pM_0,\pP_0$.
The condition  of vanishing $\jt 0 0 $ on the  physical states, namely,
$$J^{(total)}{\rM_{-n}}^s{\gM_{-n}}^r|\pM_0,\pP_0> =
(1-s-r+\pM_0){\rM_{-n}}^s{\gM_{-n}}^r|\pM_0,\pP_0> = 0\eqn\mishJs$$
gives the relation  $$s+r=\pM_0+1\eqn\mishR$$
which also agrees with the $L_0^{(total)}=0$ condition, i.e.
$ (\pM_0+1)(\pP_0-1) = -n(s+r)$.

For positive $\pP$ momentum,
 $\pP_0=n+1,\ n\geq  0$, the \co  is in the span of the states
${\cP_{-n}}^s{\bM_{-n}}^r|\pM_0,\pP_0>$
where $s=0,1$ and $r=0,1,\dots$. The $\jt 0 0 = 0$ constraint
now implies that
$$s+r=-(\pM_0+1)\eqn\mishRr$$
which again agrees with the $L_0^{(total)}=0$ condition, i.e.
$$ (\pM_0+1)(\pP_0-1) = -n(s+r).$$

When $\pP_0$ is not an integer we have a one dimensional cohomology
spanned by the vacuum $|\pP_0,\pM_0=-1>$.

The nontrivial $\Qt$-cohomology states are therefore spanned by
$$|\Phi_0>=\rM_{-n}{\gM_{-n}}^r|\pP_0=-n+1,\ n> 0,\pM_0=r\rangle $$
$$|\Phi_0>={\gM_{-n}}^r|\pP_0=-n+1,n>0,\pM_0=r-1 \rangle $$
$$|\Phi_0>=\cP_{-n}{\bM_{-n}}^r|\pP_0=n+1 > 0,\pM_0=-r-2 \rangle$$
$$|\Phi_0>={\bM_{-n}}^r|\pP_0=n+1 > 0,\pM_0=-r-1 \rangle $$
for $r=0,1,2,...$, and
$$|\Phi_0>=|any \pP_0,\pM_0=-1>.$$

We can now insert $|\Phi_0>$ into the expressions for  the
states in the \co of $\QBr$ as follows
  $$|\Psi>=  \sum_{n=0}^{\infty}{{(-1)^n} \over {n!}}R^n
           \sum_{m=0}^{\infty} {{P_0}^m \over {m!}}|\Phi_0> =
           e^{-R}e^{P_0}|\Phi_0>  $$ where
$$P_0=-\sum_{n \neq 0} {1\over n}\pP_n(\bM_m\gM_{-m-n}+\cP_m\rM_{-m-n})$$
($P_0$ is obtained from $P$  by omitting  the
operators that are zero on $|\Phi_0>$), and
$$R=\sum(\cP_n\rP_l\gM_m\gM_{-n-l-m} - \cP_n\rP_l\gP_m\gP_{-n-l-m})
   +2\sum_{n\neq 0}\cP_m\r0_n\gP_{-n-m}.$$
 This way  we obtain explicitly the cohomology of $\QBr$.

\section{ Physical states of the \SOS\ models versus those of the
$c\leq 1$ models.}

 Let us now compare the structure of the cohomology ring of
our twisted \SOS\ model to that of the $c\leq 1$ Liouville model.
At  ghost number $N_G=-1$,
we expect that the discrete states found above would
correspond to elements of the  ground ring\refmark\Wgr
(recall the shift in the ghost number when moving from states to operators
because $|0>_{phys}=\ch 1 + |0>_{SL(2,C)} $). The lowest level
state is simply $\rh {-1} - |\pP_0=\pM_0=0>$ which corresponds to the
identity operator. The next two states  of  the   \co
 of $\Qt$ which are at level 2 translate  into operators
in the \co of $\QBr$  as follows :
$$\eqalign{ \rh {-1} - \gM_{-1}|\pP_0=0,\pM_0=1>& \rightarrow \t x
=\gM e^{i\pP}\cr
  \rh {-2} - |\pP_0=-1,\pM_0=0>&\rightarrow
\t y = [-i\pa \pP + \ch {} + (\rh {} - +2 \rh {} 0 \gP + \rh {}
+[(\gM)^2 - (\gP)^2])]e^{-i\pM}\cr}\eqn\mishxy$$
 These states are (with the
 appropriate identification) at the same momenta
as those of  the ground ring generators in the $c\leq 1$ models.
  In fact $\t y$
is  equal to $y$ of ref. [\Wgr] with some additions from the ``topological
sectors". One can also change
the form of $\t x$ so it resembles that of the
ground ring $x$  by adding a $\QBr$ exact term as follows
$$\t x =\{\QBr, \half \rh {} 0 (\bM)^{-1}e^{i\pP}\} +
({\bM})^{-1}(\ch {} + \rh {} - +i\pa\pM + \bP\gP -\ch {} - \rh {} +
)e^{i\pP}\eqn\mishtx$$    The ground ring cohomology is now
generated by
 $\t x^n \t y^m$.
As in  the ground ring of ref. [\Wgr], it is easy to realize that
 area preserving diffeomorphisms leave the ground ring invariant.  These
$W_\infty$ transformations
are generated by  currents constructed by acting on
the $N_G=1$ \co operators with $G_{-1}$. Recall that $G= \rh {} - (J^+
- I^+) + 2\rh {} 0 (J^0 - I^0) + \rh {} +(J^- - I^-) +  \pa\rh {}
0$. For instance the generators $\pa_{\t x}$  and
 $\pa_{\t y}$  take the following form
$$\eqalign{\pa_{\t x}=& G_{-1}(\ch {} + e^{-i\pP}) = \bM e^{-i\pP} \cr
\pa_{\t y}=& G_{-1}(\ch {} + (\bM)^{-1} e^{i\pM} )= e^{i\pM}
\cr}\eqn\mishdxdy$$
It is easy to check that indeed, as is hinted by  the notations,
$$\pa_{\t x} \t x= \pa_{\t y} \t y= 1\qquad
\pa_{\t x} \t y= \pa_{\t y} \t x=0.\eqn\mishdx$$
 One may wonder about the operator $({\bM})^{-1}$ which
does not seem to be
an appropriate operator to use since  $\bM= J^+-I^+$.
Without the inclusion of arbitrary  powers of $\bM$ the
space of physical states of the \SOS\ model does not recover that
of the $c\leq 1$ models. As will be clarified soon a similar
situation is facing us also in the tachyonic sector. In the summary
we raise another   possible prescription for regaining
a full equivalence in the
states. Here we implement an idea of ref. [\MuVa] where a further
bosonization is invoked for  the $(\bM, \gM)$ system  as follows
$\bM\equiv e^{u-iv} $ and $\gM\equiv -i\partial ve^{-u+iv}$,
where $u,v$ are free
bosons with a background charge of $-\half$ and ${i\over 2}$
respectively. In terms of the latter bosons, one is entitled to
take any arbitrary power of $\bM$ and hence we complete the missing
states in the comparison with the gravitational models.

One branch of the tachyons of the $c\leq 1$ model can be easily
identified with a sector of the \co of the \SOS\ model: this is the
vacuum of the latter, $|\pP_0=p^+,\pM_0=-1>$ which corresponds to the
operator $\cP e^{ip^+\pM -i\pP}$. If one identifies $\phi_J$ with
the matter field $X$, $\phi_I$ with the Liouville field $\phi$
and $\cP$ with $c$, the tachyonic states of one branch are indeed
found.  However,  the other branch $\cP e^{ip^-\pP +i\pM}$,
is missing in the \co of $\QBr$.  There are, however,
additional states  with no excitations at $N_G=0$. These are the
states $(\bM_0)^r |\pP_0=1,\pM_0 =-r-1>$ corresponding to the
operators $\cP (\bM)^r e^{i\pM_0\pP +i\pM}$.
Apart from the appearence of the operator
$\bM$ these states are identical  to a discrete series of the other
branch of the tachyons.  If again we bosonize $\bM$ then $r$ can
take  any real number  and thus one finds states which
correspond to the full missing branch.
For $k=-1$, restricting the
values of $r$  to the integers would correspond to
the $c=1$ model at the self-dual radius.

The states of other ghost number are also in one to one
correspondence with those of the $c\leq 1$ Fock space relative
cohomology. The only exception is that our second branch of the
tachyon appears in both $N_G=1$ and $N_G=2$ whereas in the Liouville
model it appears only in the former.  A similar  situation was
revealed in the \SOU\ analysis of  ref. [\MuVa].
\section{ The twisted ${SL(2,R)\over U(1)}$ case}
 The  \tGH model\refmark\Wgg\ is  a
 twisted $N=2$ supersymmetric
 $G$-WZW model of level $k$  coupled to gauge
fields in the algebra of $H\in G$.  It is thus  the usual
\GH model with an additional  set of $(1,0)$ anti-commuting ghosts
which  take their values in the (negative, positive) roots of
\GH respectively. The derivation of the quantum action and the
analysis of the algebraic structure  of these models were  presented
in refs. [\NaSu,\usss].  In particular a Kac-Moody algebra
at level zero  associated
with the group $H$  was identified and used to further
twist the model in a similar manner to the \G case, namely
$T\rightarrow \t T=T+\pa \sum_{i\in CSA}(J^{(tot)})^i $,
where the summation
is  over the Cartan  sub-algebra.
 Using the $(+,-)$ bosonization
the  physical states in  the free Fock space,  as well as in  the space of
 irreducible representations, were found  in ref. [\usss]
 by computing the \co
of $Q$, the sum of the BRST  gauge-fixing charge and the twisted
supersymmetry charge.
An elaborate analysis of the \SOU\
case  at level $k=-3$ was recently  given in
ref. [\MuVa].  Here we briefly describe the application of the
method used above for the \G case to that of \SOU.
We continue to   parametrize the $J$ sector as in eqn.\mishbos,
whereas from   the $I$  sector only $I^0= \sqh\dpI$ is left
over.  This implies that  now we have $\bM=\bP=\bJ$, and
$\gM=\gP=\half\gJ$.
 The other differences relative to the content of
the \SOS\ model are the absence of the $\cM,\rP$ pair and the
change of the   pre-factor  in front of the $\rh {} 0, \ch {}
0$, for instance  in eqns. \mishOP,\mishJ,  from 2 to  1.
     Introducing  all these
alterations to the derivation of
$\Qt$  one finally finds an expression which
differs from eqn. \mishQtr\ by the omission of  the $\cM\bP$ term .
Obviously, this  also implies the
absence of the second term in the sum of Fock
spaces  in eqn.\mishFo .   Since this term was anyhow irrelevant in
the determination of the non-trivial cohomology, it follows that the
space of physical states of the \SOU\
model is identical to that of the \SOS\
model.  In particular  the $c=1$ Liouville model, that was shown to
have the same physical  states as the $k=-1$ \SOS\ model, can now be
associated with either a $k=-3$ \SOU\ model if the $J$ sector corresponds
to the gravity sector, or a $k=-1$
\SOU\ model if the $J$ sector is related to
the matter sector. Notice that in
the former case $\bM=\bJ$ is related to the
Liouville sector of the model and thus assigning it to $\sqrt{\mu}$
($\mu$ here is the cosmological constant), as was
suggested in ref. [\MuVa], seems more natural than in the \SOS\ model.
  The equivalence of the cohomologies of the \SOS\ and
\SOU\ was revealed also
  in the $(+,-)$ bosonization scheme,\refmark\usss and recently
in the appendix of ref. [\MuVa].
\section{ Summary and Discussion}
In this note we have constructed the \co of  the BRST
charge of the twisted  \SOS\ models at an arbitrary level $k$. The space
of physical states was expressed on a Fock space of free fields.
The corresponding operators were also written down. As a by product
we have identified the \co of the twisted ${SL(2,R)\over
U(1)}$, as well. At level $k=-1$ we have demonstrated that all the
states of the $c=1$ Liouville theory  were recovered. In the
general case of $k={p\over q} -2$   if one further reduces the \co
 to a Kac-Moody irreducible representation,\refmark\BF\
 a correspondence to $(p,q)$
minimal models coupled to 2D gravity is revealed.\refmark\us\
It is important to note that on the level of  the Fock space,
 all the models of  different $k$ are essentially the same, as
is depicted by the fact that $t=k+2$ could in fact be absorbed
into the definitions of the free fields $\pP$ and $\pM$.
Moreover,  the  twisted ${SL(2,R)\over U(1)}$  model also produces
the same cohomology on the Fock space.
An area preserving diffeomorphism  associated with a $W_\infty$
algebra could be identified ( as long as we stay on the Fock space
). This is the situation for the $c=1$ case where no further Felder
reduction is needed. For $c<1$ the screening charges do not commute
in general with the $W_\infty$ currents.
A similar analysis of
general \G and \GH models is under current investigation.

In order to account for all the states of the $c=1$ gravitational
model it was not sufficient to work in the $\beta^\pm, \gamma^\pm,
\phi^\pm$ Fock space associated with the Wakimoto representation
of the  Kac-Moody currents. Following ref. [\MuVa] we had to
consider states/operators involving also $(\bM)^{-r}$.  Negative
(and, in fact, arbitrary real)
powers of $\bM$ could make sense if we  further bosonize the
$\beta,\gamma$ system. Had
we limited ourselves to the $\beta^\pm, \gamma^\pm,
\phi^\pm$ fields  using the  $(+,+)$ bosonization,  we would have
recovered the ground ring generators $x,y$
(and all other discrete states)
as well as one tachyonic
branch $T^+$ and discrete states from the other branch.
 If instead we use the $(+,-)$ bosonization, ( as was
used in ref. [\us,\uss,\usss]), we obtain only the other tachyonic
branch $T^-$. The union of these two Wakimoto\refmark\Wak\
representations produces all the physical states. However, we do not
have any good reason why both representations should be considered
simultaneously. We also note that the transformation
$\pP\rightarrow -\pM$,  which amounts to $t\rightarrow {1\over t}$
and    $\phi_J\rightarrow -\phi_J$, leads to a \co composed of the
discrete states
and $T^-$. Recall that the analogous transformation in $c(t)$
matter theory coupled to gravity, $t\rightarrow {1\over t}$ and
$X\rightarrow -X$ ,  is a symmetry of the energy momentum tensor.

To establish the full isomorphism between the string theory and the
corresponding  topological model we have to show that the
correlation functions match as well. It is quite important to
establish this relation also from a practical point of view. It may
allow us to develop and use topological tools for calculations of
string correlation functions.
 Such calculations were done for the \SOU\ case in ref. [\Wgga].
 A step towards establishing such a
correspondence was made very recently in ref. [\MuVa] relating the string
theory and the topological  ${SL(2,R)\over U(1)}$  model.
It was argued, and explicitly demonstrated for the 4-tachyon amplitudes,
that the latter model yields the correlators of the $c=1$ string theory
at non-zero cosmological constant.
It is important to perform calculations
at higher genus to establish whether
one needs to further couple the topological
 \SOU\ model to topological gravity.\refmark\Wgga\
 The cosmological constant was
identified in ref. [\MuVa] with $\beta$.
Note that this identification makes the appearance
of negative powers of $\bM$ more acceptable. Similar arguments can be
put forward for the  \SOS\  model.
Note that for this case $\bI$ has to be identified
with $-\sqrt{\mu}$ while $\bJ$ should
be set to zero due to the matter momentum
conservation, namely, there is no need
for a screening charge in this sector.
  This is in accordance with the
argument of ref. [\BLNW] that the topological \G models and the standard
continuum description of $W$-gravity yield string theories at different
values of the cosmological constant.
The real   challenging question is whether the duality between string
theories and topological theories always holds, and in particular whether
(super) string theories admit a TFT  description.

 \ack{ We would like to thank E. Kiritsis and N. Marcus for useful
conversations.}

\appendix { Explicit Derivation of the Transformed Cohomology}

We start by splitting $\QBr$  into two terms  which are
distinguished by their $(\cP,\rM)$-ghost number as follows :
 $$\QBr=Q_1 + Q_2, \eqn\mishQB$$
where
$$\eqalign{Q_1=\intz&[\cM\bP+2i\c0\dpM
+2\c0(\bP\gP+\bM\gM+\cP\rM-\cM\rP)]\cr
Q_2=\intz&[\cP(4\d\gP-2t\d\gM-\bP\gP\gP-2\bM\gM\gP-\bP\gM\gM \cr
&-2i\gP\dpM-2it\gM\dpP+2\cM\r0)].\cr}\eqn\mishQQ$$
The zero modes $\r0_0$  and $\ch 0 0 $
are understood to be omitted in the  following  mode expansions.
We start by noting that
$Q_2=[Q_1,R]+\widetilde{Q_2}$
with
$$R=\intz(\cP\rP\gM\gM + 2\cP\r0\gP - \cP\rP\gP\gP)$$
$$\widetilde{Q_2} = \intz(-2t\cP\d\gM -2it\cP\gM\dpP).$$

It can be checked that:
$[\widetilde{Q_2},R]=0$ and $[[Q_1,R],R]=0$.
Thus,
$$e^{-R} Q_1 e^R = Q_1 + [Q_1,R] + \half[[Q_1,R],R] +\dots
                  = Q_1+Q_2-\widetilde{Q_2} = \QB - \widetilde{Q_2}$$
and so
$$e^R \QB e^{-R} = Q_1 + \widetilde{Q_2}.$$

We will now further decompose
$\widetilde{Q_2} = \widetilde{Q_2}^\prime+\widetilde{Q_2}^{\prime\prime}$
where
$$\widetilde{Q_2}^\prime=-2t\intz(\cP\d\gM+\pP_0\cP\gM)\qquad
\widetilde{Q_2}^{\prime\prime}=-2it\intz(\cP\gM\dpP)$$
(the rhs of the second expression
does not include the zero mode $\pP_0$),
and,
$$Q_1={Q_1}^\prime+{Q_1}^{\prime\prime}$$
$${Q_1}^\prime=\intz(\cM\bP+2i\c0\dpM)$$
$${Q_1}^{\prime\prime}=\intz(2\c0(\bP\gP+\bM\gM+\cP\rM-\cM\rP)).$$
We now define the zero dimension zero ghost number operator $P$ by
$$ P=-\intz(i\pP(\bP\gP+\bM\gM+\cP\rM-\cM\rP))\eqn\mishPR$$
where it should be
 understood that in $\pP$ we do not include the terms
$\widetilde{\phi}^+ + \pP_0 log(z)$
($\widetilde{\phi}^+$ being canonically
conjugate to $\pM_0$). In other words in mode expansion $P$
takes the form
$$P=-\sum_{n \neq 0} {1\over n}\pP_n(\bP_m\gP_{-m-n}
                     +\bM_m\gM_{-m-n}+\cP_m\rM_{-m-n}-\cM_m\rP_{-m-n}).$$

Now we can calculate:
$$\eqalign{{Q_1}^{\prime\prime}&=-[Q_1^\prime,P]\qquad
[{Q_1}^{\prime\prime},P]=-[[Q_1^\prime,P],P]=0 \cr
[\widetilde{Q_2}^\prime,P]&=-\widetilde{Q_2}^{\prime\prime}\qquad\qquad
[\widetilde{Q_2}^{\prime\prime},P]=0\cr}$$

So,
$$e^P \widetilde{Q_2}^\prime e^{-P}
  =\widetilde{Q_2}^\prime+\widetilde{Q_2}^{\prime\prime}=\widetilde{Q_2}$$
$$e^P {Q_1}^\prime e^{-P}
  ={Q_1}^\prime+{Q_1}^{\prime\prime}=Q_1$$
and finally,
$$e^{-P}e^R \QB e^{-R}e^P = {Q_1}^\prime + \widetilde{Q_2}^\prime =
Q_{tr}$$
 The cohomology of $Q_{BRST}$ is, therefore,
 isomorphic to that of $Q_{tr}$
(defined by the above equation)
which is much easier to compute.
\refout
\end